\documentclass[conference]{IEEEtran}
\IEEEoverridecommandlockouts
\usepackage{cite}
\usepackage{amsmath,amssymb,amsfonts}
\usepackage{algorithmic}
\usepackage{graphicx}
\usepackage{textcomp}
\usepackage{xcolor}
\usepackage{siunitx}
\def\BibTeX{{\rm B\kern-.05em{\sc i\kern-.025em b}\kern-.08em
    T\kern-.1667em\lower.7ex\hbox{E}\kern-.125emX}}
\begin{document}

\title{6G Infrastructures for Edge AI: An Analytical Perspective\\
}

\author{\IEEEauthorblockN{ Kurt Horvath}
\IEEEauthorblockA{\textit{Department of Information Technology} \\
\textit{University of Klagenfurt}\\
Klagenfurt, Austria \\
kurt.horvath@aau.at}
\and
\IEEEauthorblockN{ Shpresa Tuda}
\IEEEauthorblockA{\textit{Faculty of Information Sciences} \\
\textit{Mother Teresa University}\\
Skopje, North Macedonia \\
shpresa.tuda@students.unt.edu.mk}
\and
\IEEEauthorblockN{ Blerta Idrizi}
\IEEEauthorblockA{\textit{Faculty of Information Sciences} \\
\textit{Mother Teresa University}\\
Skopje, North Macedonia\\
bi211101.student@unt.edu.mk}
\and
\IEEEauthorblockN{ Stojan Kitanov}
\IEEEauthorblockA{\textit{Faculty of Information Sciences} \\
\textit{Mother Teresa University}\\
Skopje, North Macedonia \\
stojan.kitanov@unt.edu.mk}
\and
\IEEEauthorblockN{ Fisnik Doko}
\IEEEauthorblockA{\textit{Faculty of Information Sciences} \\
\textit{Mother Teresa University}\\
Skopje, North Macedonia \\
fisnik.doko@unt.edu.mk}
\and
\IEEEauthorblockN{ Dragi Kimovski}
\IEEEauthorblockA{\textit{Department of Information Technology} \\
\textit{University of Klagenfurt}\\
Klagenfurt, Austria \\
dragi.kimovski@aau.at}
}

\maketitle

\begin{abstract}
The convergence of Artificial Intelligence (AI) and the Internet of Things has accelerated the development of distributed, network-sensitive applications, necessitating ultra-low latency, high throughput, and real-time processing capabilities. While 5G networks represent a significant technological milestone, their ability to support AI-driven edge applications remains constrained by performance gaps observed in real-world deployments. This paper addresses these limitations and highlights critical advancements needed to realize a robust and scalable 6G ecosystem optimized for AI applications. Furthermore, we conduct an empirical evaluation of 5G network infrastructure in central Europe, with latency measurements ranging from \qty{61} {\milli\second} to \qty{110} {\milli\second} across different close geographical areas. These values exceed the requirements of latency-critical AI applications by approximately 270\%, revealing significant shortcomings in current deployments. Building on these findings, we propose a set of recommendations to bridge the gap between existing 5G performance and the requirements of next-generation AI applications.
\end{abstract}


\section{Introduction}
The proliferation of distributed, network-sensitive applications has surged dramatically in recent years, fueled by the convergence of Artificial Intelligence (AI) and the Internet of Things (IoT) \cite{kimovski2021cloud}. These advancements have relied heavily on mobile network technologies, with 5G representing the most recent milestone. However, as we approach the era of 6G, it is crucial to evaluate the real-world performance of 5G networks and their ability to support AI-driven edge applications effectively.

AI-enabled applications, such as autonomous vehicles, immersive virtual and augmented reality, robotics, and IoT systems, demand ultra-low latency, high throughput, and real-time data processing \cite{Siddiqi20195GDevices}. These requirements are critical for ensuring seamless operation and responsiveness. While 5G networks promise advancements like enhanced mobile broadband and ultra-reliable low-latency communication, practical deployment scenarios reveal performance gaps that hinder these applications' potential.

A primary issue limiting the deployment of AI-driven edge applications is the inability of current 5G networks to consistently deliver the low-latency performance required. Although 5G technologies claim to provide latencies as low as 1-4 milliseconds, real-world benchmarks indicate significant discrepancies. For example, tests conducted in central Europe, including a 5G network in Klagenfurt, measured latencies ranging from 7 to 12 milliseconds when connecting to the \textit{Exoscale Cloud} \cite{horvath24sealcc}. These values fall short of the stringent thresholds for latency-critical applications like autonomous vehicle coordination or real-time robotics control.

Additionally, current 5G networks lack the robust integration of cloud and edge computing resources necessary to support scalable AI applications \cite{Xiao2015GISAlgorithms}. Without dedicated, high-performance connections between the edge, core, and cloud, applications suffer from degraded performance, reduced scalability, and limited responsiveness. This infrastructural gap poses a significant roadblock to enabling the intelligent edge computing envisioned in the 6G era.

The main goal of this work is to bridge the gap between real-world 5G performance and the requirements of emerging AI applications. It highlights the essential advancements to enable a robust, intelligent, and scalable 6G ecosystem.  Therefore, this paper addresses the critical gaps in supporting AI-driven applications at the edge by:
\begin{itemize}
    \item Defining a set of characteristics and conducting a requirements analysis of the performance of the current communication networks for AI applications. This includes examining next-generation AI systems' latency, throughput, and scalability needed to inform the design of 6G technologies.
    \item Conducting a detailed real-world evaluation of 5G infrastructure in central Europe, including latency, bandwidth, and scalability benchmarks, to identify the specific shortcomings of 5G networks in meeting the demands of AI-based IoT applications.
    \item Providing deployment solutions for more efficient utilization of communication networks for AI services. 
\end{itemize}

The paper has six sections. Section \ref{sec:Formal} characterizes the 6G network infrastructures, while Section \ref{sec:AppReq} presents a formal application requirements analysis. Section \ref{sec:Net} presents a real-life evaluation of the currently available 5G networks and identifies their shortcomings. Section \ref{sec:Recom} defines a set of recommendations for the future 6G networks for seamless supporting of AI applications on the Edge. We conclude the paper by summarizing the contributions in Section \ref{sec:Conclusion}.

\section{6G Network Infrastructure Characteristics}
\label{sec:Formal}

This section outlines the network infrastructure characteristics for 6G networks to support AI at the edge, emphasizing their synergy and mutual benefits across key applications.

\subsection{Communication latency}
A defining characteristic of 6G is its ultra-low latency, which can reach as low as one hundred microseconds — ten times lower than 5G’s 1-millisecond latency \cite{she2020deep}. This improvement is critical for AI systems requiring real-time data processing and decision-making in applications such as robotics, remote surgery, and autonomous vehicles.

By minimizing network delays, 6G can empower edge AI systems to process data locally on devices like single-board computers or smart sensors, enabling faster and more efficient decision-making. For instance, 6G connectivity in autonomous vehicles would allow the deployment of AI models to analyze multi-modal sensor data in real-time, ensuring safe navigation in dynamic environments \cite{noor20226g}. Similarly, in remote surgery, AI-driven support systems will enable surgeons to perform intricate procedures precisely, regardless of physical distance \cite{nayak2021}.

\subsection{Network capacity}
AI’s reliance on large datasets for training and decision-making necessitates high-speed data transfer. With 6G offering data rates up to 1 terabit per second, these datasets can be transmitted and processed almost instantaneously \cite{chataut2024}. This leap in data velocity will transform sectors such as healthcare, manufacturing, and entertainment by enabling faster and more informed decision-making.

For example, AI algorithms in healthcare can analyze high-resolution imaging and genomic data in real-time, providing immediate diagnostic feedback and facilitating early detection of diseases such as cancer and cardiovascular disorders \cite{nayak2021}. In the entertainment sector, AI-driven platforms will deliver personalized content instantaneously, while real-time streaming of 4K and 8K media will ensure seamless user experiences.

\subsection{Network scalability}
With the support of hundreds of thousands of devices per square kilometer, 6G vastly outperforms 5G’s limit  \cite{maduranga2024}. The highly scalable connectivity will foster expansive IoT ecosystems in domains like smart cities, industrial automation, and precision agriculture, with AI serving as the cornerstone for managing and optimizing the real-time data generated by billions of connected devices.

In smart cities, AI can optimize energy consumption, monitor traffic patterns, and predict infrastructure maintenance needs. In precision agriculture, AI-driven analysis of data from connected sensors can enable efficient resource management and improved crop yields \cite{polymeni2023impact}. Similarly, industrial automation will leverage AI to monitor machinery, dynamically adapt manufacturing processes, and enhance productivity \cite{sheraz2024}.

\section{Application Requirements}
\label{sec:AppReq}
Building on the diverse characteristics presented above, the following sections focus on the requirements for latency, scalability, and capacity in 6G networks, addressing the challenges and opportunities they present.

\subsection{Communication requirements}
The communication requirements are mostly influenced by the latency, which can be divided into two categories: processing time and network latency. For the purpose of this work, we, therefore, focus on network latency. According to \cite{Ren22} and \cite{Liu19}, augmented reality applications demand motion-to-photon latency below 20 milliseconds to prevent motion sickness, while video applications require a minimum frame rate of 60 FPS, corresponding to a frame interval of 16.6 milliseconds. 

To meet these stringent requirements, 6G must achieve communication latencies as low as 100 microseconds \cite{she2020deep}. This improvement enables real-time applications such as robotics, remote surgery, and autonomous vehicles \cite{noor20226g}. Additionally, minimizing delays in IoT protocols like MQTT, AMQP, and CoAP, which contribute an extra 5-8 milliseconds \cite{Baylms2022AThings}, will be essential for achieving user-perceived latency below 16 milliseconds \cite{Liu19}.

\subsection{Bandwidth Requirements}
The bandwidth requirements for 6G networks are highly stringent, with a maximal data rate of up to 1 terabit per second, allowing the support of applications that generate vast amounts of data \cite{chataut2024}. 

To begin with, autonomous vehicles are expected to generate up to 4 terabytes of data daily. This data encompasses sensor inputs, high-definition mapping, and real-time environmental analysis, necessitating substantial network bandwidth for seamless operation and communication. Similarly, remote medical applications, including telemedicine and remote surgeries, will demand high data rates and low latency to transmit high-definition video, haptic feedback, and other critical medical data in real-time. The exact bandwidth requirements can vary depending on the specific application and the quality of the data being transmitted; however, it is in the range above 10 gigabytes per day.


\subsection{Scalability requirements}
The scalability requirement is a cornerstone of 6G networks, essential for supporting the explosive growth of interconnected devices and data-intensive applications. Estimates suggest that by 2030, over 125 billion devices will be connected globally, demanding unprecedented network capacities and seamless integration \cite{sheraz2024}. 

In smart cities, scalability enables real-time management of massive IoT ecosystems, including millions of sensors and cameras monitoring energy use, traffic, and infrastructure health. For example, adaptive traffic management systems in large cities like Tokyo could simultaneously analyze data from up to 50,000 intersections to reduce congestion. Similarly, energy grids powered by 6G-connected IoT devices can optimize consumption in real time across thousands of buildings, improving efficiency and sustainability.

Industrial automation also relies heavily on scalability, with smart factories deploying tens of thousands of sensors and devices that must communicate seamlessly. A fully automated manufacturing line can generate over 5 terabytes of data per day, requiring 6G networks to allocate resources to ensure real-time adjustments dynamically.

\section{Current Infrastructure Evaluation}
\label{sec:Net}
In this section, we conduct an in-depth empirical analysis of 5G network performance in central Europe to bridge the gap between theoretical and real-world performance.

\subsection{Use-case application}
In this context of AI and fast network connectivity, we explore a use case from the realm of augmented reality gaming, where stringent low-latency connectivity requirements arise among users located in different geographic areas. The application involves a distributed version of a fast-paced game, where two teams compete to eliminate their opponents by throwing virtual balls and avoiding incoming throws, all while wearing augmented reality headsets such as the Meta Quest 3. Players are eliminated if they are hit by a ball and fail to catch it mid-air. In our scenario, the balls are virtual, thrown using a remote controller, and the player must evade them based on their real-time visual perception.

The application comprises three core interacting services. The first service, the \textit{Video Streaming Service}, connects pairs of players, enhancing their vision to allow them to see their opponent's virtual ball in the augmented environment. To facilitate interaction, the second service, the \textit{Remote Controller Service}, enables the player to aim at the opponent and trigger an event within the \textit{Trajectory Service}. This service collects the event data—such as the direction and force of the aim—applies it to the incoming video stream, and visualizes the ball’s trajectory during flight.

To emulate the gameplay experience, we use the \texttt{ffmpeg} codec suite to create a bidirectional video stream between multiple locations environments. The reference node establishes a logical video stream that connects the players. The communication between the services must meet the stringent timing requirements of the video frame update cycle to ensure a smooth and accurate gameplay experience. This is crucial to prevent scenarios where a player is struck by a ball even though their physical location no longer aligns with the virtual ball’s position. As detailed in \cite{Mossel2012Artifice-augmentedCollaboration}, we define the maximum acceptable round-trip latency as \SI{20}{\milli\second}.

\subsection{Testbed and scenario}
For the case study evaluation, we deployed an AI service at the University of Klagenfurt, which acts as the reference node for the network analysis. Specifically, we used the RIPE Atlas probe hosted at the University of Klagenfurt as the reference \cite{horvath2023mesdd}. The network analysis focuses on evaluating network nodes within densely populated urban environments, specifically targeting devices equipped with wireless connectivity. This campaign employs a mobile reference node utilizing 5G technology to navigate through various geographic regions, referred to as \textit{sectors} $S$. Each sector encompasses \texttt{urban} residential areas in the vicinity of the University of Klagenfurt. To facilitate analysis, we apply geographical partitioning methodology \cite{Maeda2009UrbanSimulation}, thus dividing each sector into smaller units called cells $S_C \in S$. For temporal and geographical alignment of the data, we use the data format provided in \cite{StatistikAustria2024AbsoluteDensity} and specify the dimensions of the cells in \qty{1}{\kilo\meter}.

 \begin{figure}[t]
     \centering
     \includegraphics[width=0.45\textwidth]{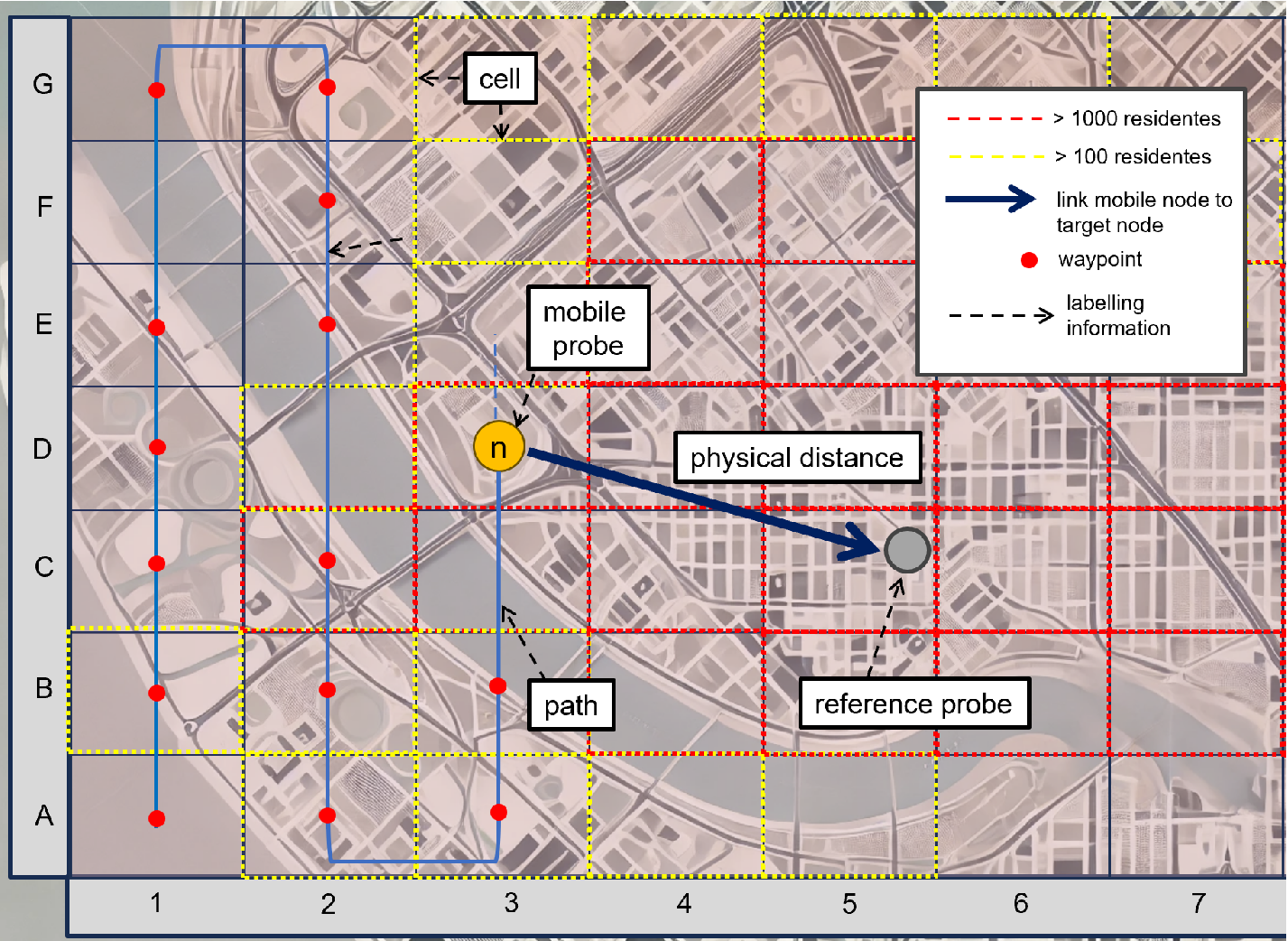} %
     \caption{Mobile evaluation scenario using grid segmentation}%
     \label{fig:scenario}%
\end{figure}

As depicted in Figure \ref{fig:scenario}, we traversed \num{33} cells (marked from A - F and 1 - 7) within a designated section of the city of \texttt{Klagenfurt}, as defined in \cite{StatistikAustria2024AbsoluteDensity}, using multiple mobile nodes. Each mobile node recorded latency measurements to eight other nodes within the same sector. The number of measurements collected per cell varied, influenced by adherence to traffic flow dynamics and local traffic regulations in Klagenfurt.

\subsection{Evaluation results}\label{sub:results}
\begin{figure}[t]
     \centering
     \includegraphics[width=0.49\textwidth]{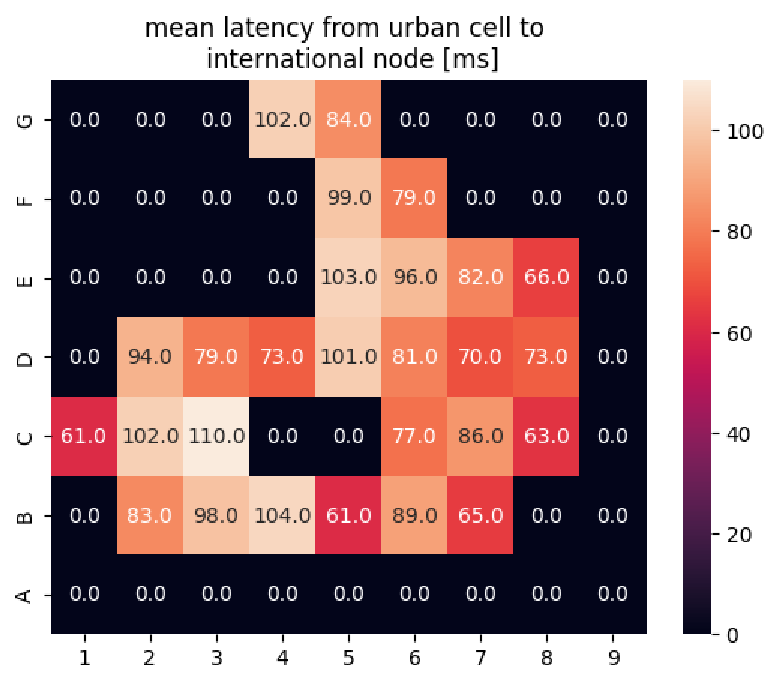} %
     \vspace{-8mm}
     \caption{Urban Mean Round-trip Time Latency}%
     \label{fig:meanlat}%
\end{figure}

As illustrated in Figure ~\ref{fig:meanlat}, the mean round-trip time latency (\texttt{RTL}) for mobile nodes surpasses that of wired nodes by a factor of seven. The \texttt{RTL} values vary significantly between cells, ranging from \qty{61}{\milli\second} at cell \texttt{C1} to \qty{110}{\milli\second} at position \texttt{C3}.  

\begin{figure}[t]
     \centering
     \includegraphics[width=0.49\textwidth]{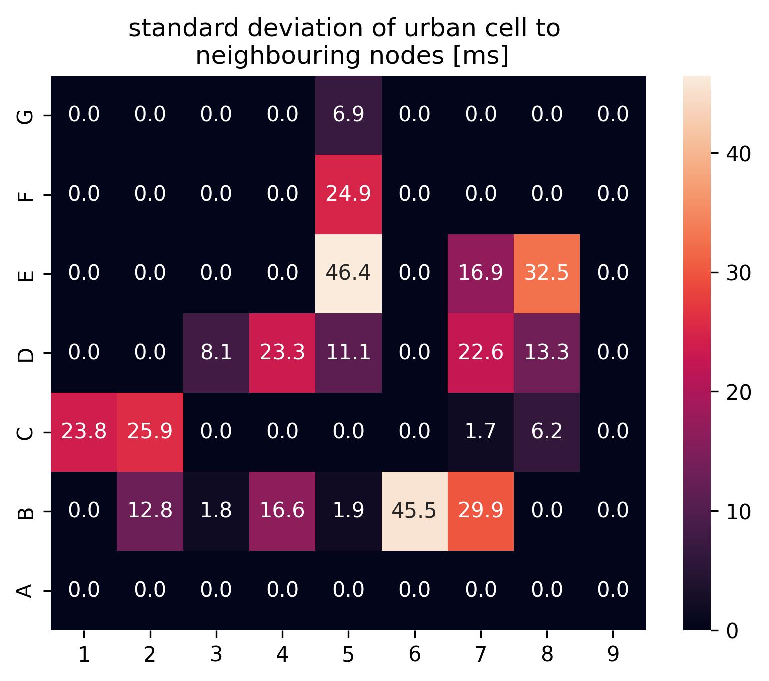} %
     \vspace{-8mm}
     \caption{Standard Deviation Latency }%
     \label{fig:stddev}%
\end{figure}

The standard deviation in this scenario, which is provided in  Figure~\ref{fig:stddev}, also demonstrates notable variability, spanning from as low as \qty{1.8}{\milli\second} in cell \texttt{B3} to as high as \qty{46.4}{\milli\second} in cell \texttt{E5}. This large variance highlights significant inter-cell and intra-cell latency differences, which are considerably higher compared to static nodes. 

Figure~\ref{fig:meanlat} further identifies a few cells marked with \texttt{0.0}, which indicate fewer than ten measurements. These instances occur primarily in border regions, where population density falls below \num{1000} inhabitants per \unit{\kilo\meter\squared}.  

The findings align with prior work, such as \cite{Shayea2021PerformanceMalaysia}, which observed similar \texttt{RTL} patterns in \texttt{urban} areas using 5G technology across various network operators. Comparable studies, such as \cite{Fan2021AnPerspective}, report latencies ranging from \qty{30}{\milli\second} to \qty{100}{\milli\second}. These results underscore the continued development of 5G and upcoming 6G technologies, aiming for sub-\qty{1}{\milli\second} latencies to achieve competitiveness with wired networks. 

Despite the relatively high standard deviation, we observe very high \texttt{RTL}, with a minimum value of 61 ms. This exceeds the identified requirements defined in Section \ref{sec:AppReq} by approximately 270\%. The discrepancy cannot be attributed solely to the 5G standard used in this evaluation. Tutti \cite{xu2022tutti} has reported similar findings \cite{xu2022tutti}, who observed an average latency exceeding 180 ms in a 5G-enabled application.
It is important to note that the results we observed only reflect the networking latency, as the RIPE Atlas network did not carry out any application-specific processing. Fezeu \cite{fezeu2023depth} evaluated the physical performance (ISO OSI networking layer 1) of 5G and reported promising results, noting that the system transmitted 4.4\% of packets in under 1 ms and 22.36\% in under 3 ms. However, the application use case significantly impacted overall latency in their evaluation. On average, the application layer added 35 ms, with the remaining latency stemming from public network delays beyond the access network and before the application's service domain.

\emph{Technical limitation of current 5G infrastructures for supporting AI applications:}
The pronounced disparities in \texttt{RTL} between mobile and static nodes in \texttt{urban} environments often exceed a fourfold difference. An analysis of data trace, derived from \texttt{RTL} measurements, reveals that the number of network hops frequently surpasses ten. This observation indicates a suboptimal 5G access network integration with the local infrastructure, likely caused by insufficient or inefficient routing pathways. Optimizing routing protocols and strengthening the interplay between mobile access networks and edge infrastructure in 6G networks is recommended to address these challenges to minimize latency for time-sensitive IoT services and support of next-generation AI applications.

\begin{figure}[t]
     \centering
     \includegraphics[width=0.45\textwidth]{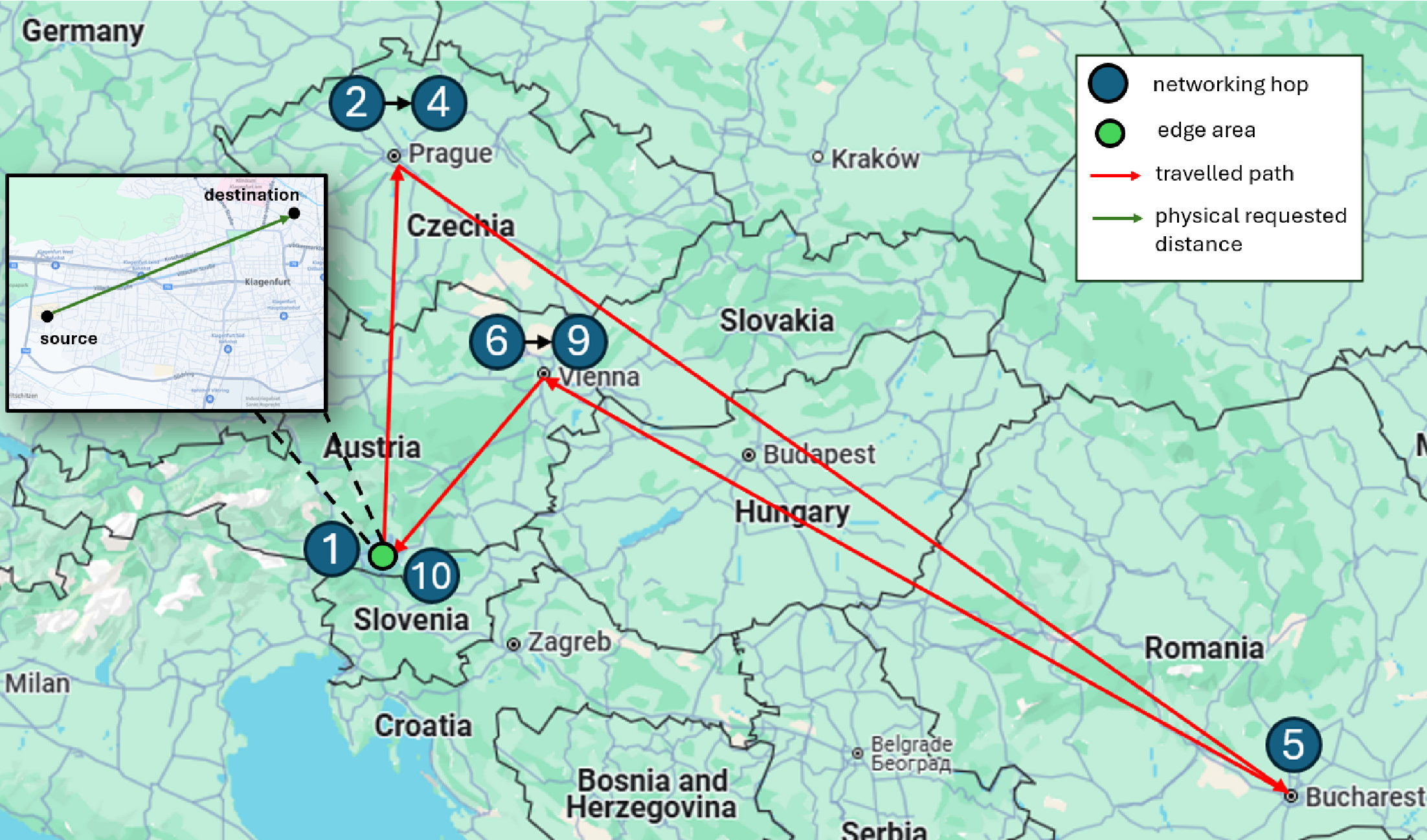} %
     \caption{Data Trace of Local Service Request }%
     \label{fig:trace}%
\end{figure}

We conducted further analysis of data trace, which is presented in Table \ref{tab:hops}, where we observed an overall \texttt{RTL} of 65 ms caused by 10 network hops. The mobile node used for this measurement is located in cell C2, while the RIPE Atlas probe is in E3, separated by less than 5 km.
Moreover, Figure \ref{fig:trace} illustrates how data transfer takes an inefficient route. On the second hop, data request moves beyond the urban area. It leaves the country, traveling from Vienna (Austria) to Prague (Czech Republic), then to Bucharest (Romania), and back to Vienna (Austria). This route covers a total distance of 2544 km. Such inefficiency undermines the goal of reducing latency through edge resources.

\begin{table}[!t]
 \renewcommand{\arraystretch}{1.3}
 \caption{Networking Hops for Local Service Request}
 \label{tab:hops}
 \centering
 \begin{tabular}{c||l}
 \hline
 \bfseries Hop & \bfseries Node\\
 \hline\hline
 1 & 10.12.128.1\\
 \hline
 2 & unn-37-19-223-61.datapacket.com {[}37.19.223.61{]}\\
 \hline
 3 & vl204.vie-itx1-core-2.cdn77.com {[}185.156.45.138{]}\\
 \hline
 4 & zetservers.peering.cz {[}185.0.20.31{]}\\
 \hline
 5 & vie-dr2-cr1.zet.net {[}103.246.249.33{]}\\
 \hline
 6 & amanet-cust.zet.net {[}185.104.63.33{]}\\
 \hline
 7 & ae2-97.mx204-1.ix.vie.at.as39912.net {[}185.211.219.155{]}\\
 \hline
 8 & 003-228-016-195.ascus.at {[}195.16.228.3{]}\\
 \hline
 9 & 180-246-016-195.ascus.at {[}195.16.246.180{]}\\
 \hline
 10 & 195.140.139.133\\
 \hline
 \end{tabular}
 \end{table}
 
\section{6G Infrastructure Recommendations for Edge AI}
\label{sec:Recom}

The deployment of low-latency Edge AI applications in 6G networks requires a multi-faceted approach to infrastructure optimization. Based on the analyses presented in Section \ref{sub:results}, we propose three key strategies: local peering optimization, User Plane Function (UPF) integration, and Control Plane Functionality (CPF) enhancement. Together, these strategies are designed to reduce latency, improve efficiency, and ensure adaptability in next-generation networks, based on our evaluation results.

\subsection{Local Peering Optimization}
\label{sub:peering}
The results in Section \ref{sub:results}, Figure \ref{fig:trace} demonstrate that routing from the source node to a local destination node involves numerous unnecessary intermediate nodes. Local peering methods eliminate these redundant paths, creating a shorter and more optimized route between the source and destination.

A significant limitation in current network architectures is the inefficient routing of data packets, which often disregards local geographic context. As documented by Castro \cite{castro2014remote} and initially identified by Huston \cite{huston1999interconnection}, this practice leads to unnecessary delays. For instance, Figure \ref{fig:trace} shows that even when a direct geographic route exists, network traffic frequently takes longer paths due to existing routing policies.

Local Internet Service Providers (ISPs) can mitigate these delays by implementing direct local peering. Research by Horvath \cite{horvath24sealcc} has demonstrated that such optimization can achieve round-trip latencies as low as \SI{1}{\milli\second}. Yet, as Figure \ref{fig:meanlat} reveals, measured latencies range from 61 to \SI{103}{\milli\second} - indicating that the majority of the delay stems from excessive networking hops rather than the physical distance traveled. Table \ref{tab:hops} further highlights that even in geographically confined regions like Prague and Vienna, multiple hops are common, exacerbating latency issues.

Numerous studies reinforce the benefits of local peering. For example, Gupta et al. \cite{gupta2014peering} showed that establishing local peering links at Internet Exchange Points reduced round-trip times within Africa from over \SI{300}{\milli\second} to as low as \SI{300}{\milli\second}. Additional solutions, such as edge peering \cite{wang24} and local caching mechanisms \cite{katsaros2017cache}, further minimize latency by streamlining data exchange between ISPs. However, despite its technical feasibility, conflicting business interests often hinder local peering adoption. Although enhanced network performance benefits all providers collectively, individual ISPs may be reluctant to collaborate due to competitive concerns. Overcoming these economic barriers is essential for fully realizing the potential of local peering in latency-sensitive 6G applications.

\subsection{User Plane Function Integration}
\label{sub:upf}
User plane function integration is essential to address the high latency as depicted in Section \ref{sub:results}, Figure \ref{fig:meanlat}. The results clearly indicate that the latency exceeds \SI{60}{\milli\second} for many instances. In contrast, Horvath \cite{horvath24sealcc} reports latencies between \SI{1}{\milli\second} and \SI{11}{\milli\second} in the same topological area using wired networks. Therefore, the choice of access network significantly impacts latency. Although the current evaluation shows 5G performing the worst, the standard includes mechanisms for ultra-low latency, such as the user plane function.\\
As a central component in both 5G and 6G architectures, the UPF facilitates edge computing and reduces reliance on core network routing. Leyva \cite{leyva2020dynamic} and Nguyen \cite{nguyen2021scaling} describe how UPFs efficiently manage AI workloads, while studies by Barrachina \cite{barrachina2024demand} and Goshi \cite{goshi2024joint} demonstrate that UPF integration can achieve latencies between 5 and \SI{6.2}{\milli\second}—a reduction of up to 90\% compared to our evaluation results exceeding \SI{62}{\milli\second} (see Figure \ref{fig:meanlat}).

Strategic placement of UPFs can eliminate excessive networking hops. Unlike the ten hops observed in our evaluation (Table \ref{tab:hops}), UPF-hosted services allow direct access by user equipment, substantially reducing latency. Moreover, the work by Jain et al. \cite{9830519} on a Smart NIC-based UPF shows that bypassing traditional host memory and PCIe bus bottlenecks can double throughput and reduce packet processing latency by a factor of 3.75 \cite{10.1145/3485983.3494861}.

The transition from 5G to 6G further accentuates the benefits of UPF integration. While 5G targeted sub-\SI{5}{\milli\second} latencies \cite{Gupta2015ATechnologies}, 6G aspires to achieve delays below \SI{1}{\milli\second} through the use of mmWave technology and advanced routing techniques. Our findings suggest that dynamic UPF selection can facilitate adaptive routing—prioritizing latency-sensitive tasks at the edge while offloading less critical workloads to centralized cloud UPFs. However, deeper UPF integration introduces the risk of vendor lock-in, as mobile ISPs' control over UPF deployment may limit third-party service provider flexibility.

\subsection{Control Plane Functionality Enhancement}\label{sub:cpf}
As discussed in Section \ref{sub:upf}, we can drastically reduce the latency initially observed in Section \ref{sub:results}. However, applications in the 5G user plane typically operate without functionalities available in the control plane, such as authentication and policy control. Since some services require this data, applying services in the control plane remains feasible.\\

Currently, a 5G Radio Access Network employs Remote Radio Heads connected to a centralized baseband Central Unit via fiber \cite{3GPPTS38401, ORAN}. In contrast, the emerging Open RAN (ORAN) paradigm offers dynamic frequency allocation, enhanced subscriber mobility management, and robust QoS enforcement via the Service Management and Orchestration framework \cite{ORAN}.

A critical shortcoming of traditional architectures is the separation of mobility management and session handling between the RAN and the core network \cite{3GPPTS23.501}. By integrating subscriber policies into the Near-Real-Time RAN Intelligent Controller (Near-RT RIC), as proposed in \cite{10073488}, it becomes possible to consolidate session and mobility management at the network edge, thereby improving decision efficiency. Nevertheless, the constraints imposed by real-time scheduling require a hybrid approach that balances centralized and decentralized control mechanisms.

Another essential aspect of control plane optimization is the strategic deployment of Network Functions and the utilization of fast Service-Based Interfaces. The context-aware QoS model introduced by \cite{9830519} dynamically prioritizes Packet Detection Rules and QoS Enforcement Rules, reducing lookup and update latencies while enabling the simultaneous prioritization of multiple flows per UE.

End-to-end network slicing \cite{9003208} is critical for allocating dedicated resources to specific applications. Although virtualization layers and network hypervisors \cite{7295561} enable such slicing, achieving real-time adaptability remains challenging. Current hypervisor placement strategies focus on latency reduction \cite{8331922}, resilience \cite{9367017}, and load balancing \cite{amjad2021delay}, yet they typically operate in a reactive rather than predictive manner. 

\section{Conclusion}
\label{sec:Conclusion}
In this paper, we analyzed the defining characteristics of future 6G communication networks and the unique requirements of AI-driven applications, with a particular emphasis on the role of edge computational resources. To bridge the gap between theoretical advancements and real-world performance, we conducted an in-depth empirical analysis of 5G network performance in central Europe, focusing on latency metrics and their implications for AI workloads.

Our evaluation revealed significant variability in round-trip latency across different network configurations and environments. For mobile nodes, the mean latency surpasses that of wired nodes by a factor of seven. These findings underscore the limitations of current 5G networks in meeting the stringent latency requirements of AI applications, which demand sub-\qty{1}{\milli\second} latencies to achieve seamless performance.

Furthermore, the latency observed exceeds the requirements identified for AI-driven applications by approximately 270\%, a substantial discrepancy attributed to public network delays and application-layer overheads. Despite these challenges, promising advances have been reported in the literature, with some 5G systems achieving latencies below \qty{3}{\milli\second} under specific conditions. However, the majority of current deployments fall short of these benchmarks.


By correlating the identified requirements of AI applications with the observed capabilities of existing networks, we proposed a robust set of recommendations to guide the design and deployment of next-generation 6G networks. These recommendations emphasize the need for scalable, low-latency, and high-throughput architectures that can seamlessly integrate AI workloads within edge computing infrastructures.

Moving forward, our future work will expand the geographical scope of the evaluation to include diverse regions, environments, and network conditions. This will enable us to refine our findings and validate the proposed recommendations further. Additionally, we plan to explore emerging technologies, such as intelligent network slicing, federated learning at the edge, and energy-efficient network management, to provide a holistic framework for the evolution of 6G networks optimized for AI applications.

\section*{Acknowledgment}
This work received funding from the OeAD agency of the Austrian Ministry for Education, Science and Research and the Macedonian Ministry for Education and Science under project number MK10/2024.


\bibliographystyle{IEEEtran}
\bibliography{bib}

\end{document}